\pdfoutput=1
\documentclass[12pt,a4paper,titlepage]{article}
\usepackage[utf8]{inputenc}
\usepackage[top=50pt,bottom=50pt,left=68pt,right=66pt]{geometry}
\usepackage{amsmath}
\usepackage{booktabs}
\usepackage{amssymb}
\usepackage[title]{appendix}
\usepackage{mathrsfs}
\usepackage{float}
\usepackage{multirow}
\usepackage{graphicx,caption,subcaption}
\usepackage[space]{grffile}
\usepackage{bm}

\begin{document}
\renewcommand{\arraystretch}{1.3}

\makeatletter
\def\@hangfrom#1{\setbox\@tempboxa\hbox{{#1}}%
      \hangindent 0pt%\wd\@tempboxa
      \noindent\box\@tempboxa}
\makeatother

% Underline for text or math

\def\un#1{\relax\ifmmode\@@underline#1\else
        $\@@underline{\hbox{#1}}$\relax\fi}

% Accents and foreign (in text):

\let\under=\unt                 % bar-under (but see \un above)
\let\ced=\ce                    % cedilla
\let\du=\du                     % dot-under
\let\um=\Hu                     % Hungarian umlaut
\let\sll=\lp                    % slashed (suppressed) l (Polish)
\let\Sll=\Lp                    % " L
\let\slo=\os                    % slashed o (Scandinavian)
\let\Slo=\Os                    % " O
\let\tie=\ta                    % tie-after (semicircle connecting two letters)
\let\br=\ub                     % breve
                % Also: \`        grave
                %       \'        acute
                %       \v        hacek (check)
                %       \^        circumflex (hat)
                %       \~        tilde (squiggle)
                %       \=        macron (bar-over)
                %       \.        dot (over)
                %       \"        umlaut (dieresis)
                %       \aa \AA   A-with-circle (Scandinavian)
                %       \ae \AE   ligature (Latin & Scandinavian)
                %       \oe \OE   " (French)
                %       \ss       es-zet (German sharp s)
                %       \$  \#  \&  \%  \pounds  {\it\&}  \dots

% Abbreviations for Greek letters

\def\a{\alpha}
\def\b{\beta}
\def\c{\chi}
\def\d{\delta}
\def\e{\epsilon}
\def\f{\phi}
\def\g{\gamma}
\def\h{\eta}
\def\i{\iota}
\def\j{\psi}
\def\k{\kappa}
\def\l{\lambda}
\def\m{\mu}
\def\n{\nu}
\def\o{\omega}
\def\p{\pi}
\def\q{\theta}
\def\r{\rho}
\def\s{\sigma}
\def\t{\tau}
\def\u{\upsilon}
\def\x{\xi}
\def\z{\zeta}
\def\D{\Delta}
\def\F{\Phi}
\def\G{\Gamma}
\def\J{\Psi}
\def\L{\Lambda}
\def\O{\Omega}
\def\P{\Pi}
\def\Q{\Theta}
\def\S{\Sigma}
\def\U{\Upsilon}
\def\X{\Xi}

% Varletters

\def\ve{\varepsilon}
\def\vf{\varphi}
\def\vr{\varrho}
\def\vs{\varsigma}
\def\vq{\vartheta}

% Calligraphic letters

\def\ca{{\cal A}}
\def\cb{{\cal B}}
\def\cc{{\cal C}}
\def\cd{{\cal D}}
\def\ce{{\cal E}}
\def\cf{{\cal F}}
\def\cg{{\cal G}}
\def\ch{{\cal H}}
\def\ci{{\cal I}}
\def\cj{{\cal J}}
\def\ck{{\cal K}}
\def\cl{{\cal L}}
\def\cm{{\cal M}}
\def\cn{{\cal N}}
\def\co{{\cal O}}
\def\cp{{\cal P}}
\def\cq{{\cal Q}}
\def\car{{\cal R}}
\def\cs{{\cal S}}
\def\ct{{\cal T}}
\def\cu{{\cal U}}
\def\cv{{\cal V}}
\def\cw{{\cal W}}
\def\cx{{\cal X}}
\def\cy{{\cal Y}}
\def\cz{{\cal Z}}

% Fonts

\def\Sc#1{{\hbox{\sc #1}}}      % script for single characters in equations
\def\Sf#1{{\hbox{\sf #1}}}      % sans serif for single characters in equations

                        % Also:  \rm      Roman (default for text)
                        %        \bf      boldface
                        %        \it      italic
                        %        \mit     math italic (default for equations)
                        %        \sl      slanted
                        %        \em      emphatic
                        %        \tt      typewriter
                        % and sizes:    \tiny
                        %               \scriptsize
                        %               \footnotesize
                        %               \small
                        %               \normalsize
                        %               \large
                        %               \Large
                        %               \LARGE
                        %               \huge
                        %               \Huge

% Math symbols

\def\slpa{\slash{\pa}}                            % slashed partial derivative
\def\slin{\SLLash{\in}}                                   % slashed in-sign
\def\bo{{\raise-.3ex\hbox{\large$\Box$}}}               % D'Alembertian
\def\cbo{\Sc [}                                         % curly "
\def\pa{\partial}                                       % curly d
\def\de{\nabla}                                         % del
\def\dell{\bigtriangledown}                             % hi ho the dairy-o
\def\su{\sum}                                           % summation
\def\pr{\prod}                                          % product
\def\iff{\leftrightarrow}                               % <-->
\def\conj{{\hbox{\large *}}}                            % complex conjugate
\def\ltap{\raisebox{-.4ex}{\rlap{$\sim$}} \raisebox{.4ex}{$<$}}   % < or ~
\def\gtap{\raisebox{-.4ex}{\rlap{$\sim$}} \raisebox{.4ex}{$>$}}   % > or ~
\def\TH{{\raise.2ex\hbox{$\displaystyle \bigodot$}\mskip-4.7mu \llap H \;}}
\def\face{{\raise.2ex\hbox{$\displaystyle \bigodot$}\mskip-2.2mu \llap {$\ddot
        \smile$}}}                                      % happy face
\def\dg{\sp\dagger}                                     % hermitian conjugate
\def\ddg{\sp\ddagger}                                   % double dagger
                        % Also:  \int  \oint              integral, contour
                        %        \hbar                    h bar
                        %        \infty                   infinity
                        %        \sqrt                    square root
                        %        \pm  \mp                 plus or minus
                        %        \cdot  \cdots            centered dot(s)
                        %        \oplus  \otimes          group theory
                        %        \equiv                   equivalence
                        %        \sim                     ~
                        %        \approx                  approximately =
                        %        \propto                  funny alpha
                        %        \ne                      not =
                        %        \le \ge                  < or = , > or =
                        %        \{  \}                   braces
                        %        \to  \gets               -> , <-
                        % and spaces:  \,  \:  \;  \quad  \qquad
                        %              \!                 (negative)

\font\tenex=cmex10 scaled 1200

% Math stuff with one argument

\def\sp#1{{}^{#1}}                              % superscript (unaligned)
\def\sb#1{{}_{#1}}                              % sub"
\def\oldsl#1{\rlap/#1}                          % poor slash
\def\slash#1{\rlap{\hbox{$\mskip 1 mu /$}}#1}      % good slash for lower case
\def\Slash#1{\rlap{\hbox{$\mskip 3 mu /$}}#1}      % " upper
\def\SLash#1{\rlap{\hbox{$\mskip 4.5 mu /$}}#1}    % " fat stuff (e.g., M)
\def\SLLash#1{\rlap{\hbox{$\mskip 6 mu /$}}#1}      % slash for no-in sign
\def\PMMM#1{\rlap{\hbox{$\mskip 2 mu | $}}#1}   %
\def\PMM#1{\rlap{\hbox{$\mskip 4 mu ~ \mid $}}#1}       %
\def\Tilde#1{\widetilde{#1}}                    % big tilde
\def\Hat#1{\widehat{#1}}                        % big hat
\def\Bar#1{\overline{#1}}                       % big bar
\def\sbar#1{\stackrel{*}{\Bar{#1}}}             % big bar with star
\def\bra#1{\left\langle #1\right|}              % < |
\def\ket#1{\left| #1\right\rangle}              % | >
\def\VEV#1{\left\langle #1\right\rangle}        % < >
\def\abs#1{\left| #1\right|}                    % | |
\def\leftrightarrowfill{$\mathsurround=0pt \mathord\leftarrow \mkern-6mu
        \cleaders\hbox{$\mkern-2mu \mathord- \mkern-2mu$}\hfill
        \mkern-6mu \mathord\rightarrow$}
\def\dvec#1{\vbox{\ialign{##\crcr
        \leftrightarrowfill\crcr\noalign{\kern-1pt\nointerlineskip}
        $\hfil\displaystyle{#1}\hfil$\crcr}}}           % <--> accent
\def\dt#1{{\buildrel {\hbox{\LARGE .}} \over {#1}}}     % dot-over for sp/sb
\def\dtt#1{{\buildrel \bullet \over {#1}}}              % alternate "
\def\der#1{{\pa \over \pa {#1}}}                % partial derivative
\def\fder#1{{\d \over \d {#1}}}                 % functional derivative
                % Also math accents:    \bar
                %                       \check
                %                       \hat
                %                       \tilde
                %                       \acute
                %                       \grave
                %                       \breve
                %                       \dot    (over)
                %                       \ddot   (umlaut)
                %                       \vec    (vector)

% Math stuff with more than one argument

\def\frac#1#2{{\textstyle{#1\over\vphantom2\smash{\raise.20ex
        \hbox{$\scriptstyle{#2}$}}}}}                   % fraction
\def\half{\frac12}                                        % 1/2
\def\sfrac#1#2{{\vphantom1\smash{\lower.5ex\hbox{\small$#1$}}\over
        \vphantom1\smash{\raise.4ex\hbox{\small$#2$}}}} % alternate fraction
\def\bfrac#1#2{{\vphantom1\smash{\lower.5ex\hbox{$#1$}}\over
        \vphantom1\smash{\raise.3ex\hbox{$#2$}}}}       % "
\def\afrac#1#2{{\vphantom1\smash{\lower.5ex\hbox{$#1$}}\over#2}}    % "
\def\partder#1#2{{\partial #1\over\partial #2}}   % partial derivative of
\def\parvar#1#2{{\d #1\over \d #2}}               % variation of
\def\secder#1#2#3{{\partial^2 #1\over\partial #2 \partial #3}}  % second "
\def\on#1#2{\mathop{\null#2}\limits^{#1}}               % arbitrary accent
\def\bvec#1{\on\leftarrow{#1}}                  % backward vector accent
\def\oover#1{\on\circ{#1}}                              % circle accent

\def\[{\lfloor{\hskip 0.35pt}\!\!\!\lceil}
\def\]{\rfloor{\hskip 0.35pt}\!\!\!\rceil}
\def\Lag{{\cal L}}
\def\du#1#2{_{#1}{}^{#2}}
\def\ud#1#2{^{#1}{}_{#2}}
\def\dud#1#2#3{_{#1}{}^{#2}{}_{#3}}
\def\udu#1#2#3{^{#1}{}_{#2}{}^{#3}}
\def\calD{{\cal D}}
\def\calM{{\cal M}}

\def\szet{{${\scriptstyle \b}$}}
\def\ulA{{\un A}}
\def\ulM{{\underline M}}
\def\cdm{{\Sc D}_{--}}
\def\cdp{{\Sc D}_{++}}
\def\vTheta{\check\Theta}
\def\fracm#1#2{\hbox{\large{${\frac{{#1}}{{#2}}}$}}}
\def\ha{{\fracmm12}}
\def\tr{{\rm tr}}
\def\Tr{{\rm Tr}}
\def\itrema{$\ddot{\scriptstyle 1}$}
\def\ula{{\underline a}} \def\ulb{{\underline b}} \def\ulc{{\underline c}}
\def\uld{{\underline d}} \def\ule{{\underline e}} \def\ulf{{\underline f}}
\def\ulg{{\underline g}}
\def\items#1{\\ \item{[#1]}}
\def\ul{\underline}
\def\un{\underline}
\def\fracmm#1#2{{{#1}\over{#2}}}
\def\footnotew#1{\footnote{\hsize=6.5in {#1}}}
\def\low#1{{\raise -3pt\hbox{${\hskip 0.75pt}\!_{#1}$}}}

\def\Dot#1{\buildrel{_{_{\hskip 0.01in}\bullet}}\over{#1}}
\def\dt#1{\Dot{#1}}

\def\DDot#1{\buildrel{_{_{\hskip 0.01in}\bullet\bullet}}\over{#1}}
\def\ddt#1{\DDot{#1}}

\def\DDDot#1{\buildrel{_{_{\hskip 0.01in}\bullet\bullet\bullet}}\over{#1}}
\def\dddt#1{\DDDot{#1}}

\def\DDDDot#1{\buildrel{_{_{\hskip 
0.01in}\bullet\bullet\bullet\bullet}}\over{#1}}
\def\ddddt#1{\DDDDot{#1}}

\def\Tilde#1{{\widetilde{#1}}\hskip 0.015in}
\def\Hat#1{\widehat{#1}}

% Aligned equations

\newskip\humongous \humongous=0pt plus 1000pt minus 1000pt
\def\caja{\mathsurround=0pt}
\def\eqalign#1{\,\vcenter{\openup2\jot \caja
        \ialign{\strut \hfil$\displaystyle{##}$&$
        \displaystyle{{}##}$\hfil\crcr#1\crcr}}\,}
\newif\ifdtup
\def\panorama{\global\dtuptrue \openup2\jot \caja
        \everycr{\noalign{\ifdtup \global\dtupfalse
        \vskip-\lineskiplimit \vskip\normallineskiplimit
        \else \penalty\interdisplaylinepenalty \fi}}}
\def\li#1{\panorama \tabskip=\humongous                         % eqalignno
        \halign to\displaywidth{\hfil$\displaystyle{##}$
        \tabskip=0pt&$\displaystyle{{}##}$\hfil
        \tabskip=\humongous&\llap{$##$}\tabskip=0pt
        \crcr#1\crcr}}
\def\eqalignnotwo#1{\panorama \tabskip=\humongous
        \halign to\displaywidth{\hfil$\displaystyle{##}$
        \tabskip=0pt&$\displaystyle{{}##}$
        \tabskip=0pt&$\displaystyle{{}##}$\hfil
        \tabskip=\humongous&\llap{$##$}\tabskip=0pt
        \crcr#1\crcr}}

% Some defs

\def\eV{\,{\rm eV}}
\def\keV{\,{\rm keV}}
\def\MeV{\,{\rm MeV}}
\def\GeV{\,{\rm GeV}}
\def\TeV{\,{\rm TeV}}
\def\sv{\left<\sigma v\right>}
\def\({\left(}
\def\){\right)}
\def\cm{{\,\rm cm}}
\def\K{{\,\rm K}}
\def\kpc{{\,\rm kpc}}
\def\beq{\begin{equation}}
\def\eeq{\end{equation}}
\def\bea{\begin{eqnarray}}
\def\eea{\end{eqnarray}}

% New commands

\newcommand{\be}{\begin{equation}}
\newcommand{\ee}{\end{equation}}
\newcommand{\nbe}{\begin{equation*}}
\newcommand{\nee}{\end{equation*}}

\newcommand{\fr}{\frac}
\newcommand{\lb}{\label}

\thispagestyle{empty}

{\hbox to\hsize{
\vbox{\noindent September 2024 \hfill IPMU24-0002 \\
\noindent    \hfill }}}

\noindent
\vskip2.0cm
\begin{center}

{\large\bf Quantum loop corrections in the modified gravity model of \vglue.1in Starobinsky inflation with primordial black hole production}

\vglue.3in

Sultan Saburov~${}^{a}$ and Sergei V. Ketov~${}^{a,b,c}$~\footnote{The corresponding author}
\vglue.3in

${}^a$~Interdisciplinary Research Laboratory, Tomsk State University, Tomsk 634050, Russia\\
${}^b$~Department of Physics, Tokyo Metropolitan University, 1-1 Minami-ohsawa, Hachioji, \\
Tokyo 192-0397, Japan \\
${}^c$~Kavli Institute for the Physics and Mathematics of the Universe (WPI),
\\The University of Tokyo Institutes for Advanced Study,  Chiba 277-8583, Japan\\
\vglue.1in

saburov@mail.tsu.ru,  ketov@tmu.ac.jp

\end{center}

\vglue.3in

\begin{center}
{\Large\bf Abstract}  
\end{center}
\vglue.1in

\noindent  A modified gravity model of Starobinsky inflation and primordial black hole production was proposed in good (within $1\sigma$) agreement with current measurements of the cosmic microwave background radiation. The model is an extension of the singularity-free Appleby-Battye-Starobinsky model by the $R^4$-term with different values of the parameters whose fine-tuning leads to efficient production of primordial black holes on smaller scales with the asteroid-size masses between $10^{16}$ g and $10^{20}$ g. Those primordial black holes may be part (or the whole) of the current dark matter, while the proposed model can be confirmed or falsified by detection or absence of the induced gravitational waves with the frequencies in the $10^{-2}$ Hz range. The relative size of quantum (loop) corrections to the power spectrum of scalar perturbations in the model is found to be of the order $10^{-3}$ or less, so that the model is not ruled out by the quantum corrections.

\newpage

\section{Introduction}

The Starobinsky model of inflation \cite{Starobinsky:1980te} as the modified $(R+\alpha R^2)$ gravity is theoretically well motivated (see e.g., Refs.~\cite{Ketov:2019toi,Ivanov:2021chn} for a recent review), being in excellent agreement  with the current cosmic microwave background (CMB) radiation measurements \cite{Planck:2018jri,BICEP:2021xfz,Tristram:2021tvh}. The Starobinsky model can be extended within modified $F(R)$ gravity in order to describe double slow-roll (SR) inflation with an ultra-slow-roll (USR) phase by engineering the function $F(R)$ leading to a near-inflection point in the inflaton potential below the scale of inflation \cite{Frolovsky:2022ewg,Saburov:2023buy,Kamenshchik:2024kay}. It results in large density perturbations whose gravitational collapse leads to production  of primordial black holes (PBH).

Adding the near-inflection point and the USR phase requires fine-tuning of the model parameters \cite{Geller:2022nkr,Cole:2023wyx}, which often lowers the value of the CMB tilt $n_s$ of scalar perturbations and thus leads to a tension with CMB measurements \cite{Karam:2022nym}, while large perturbations may imply significant nonGaussianity and quantum (loop) corrections that may invalidate classical single-field models of inflation and PBH production \cite{Kristiano:2022maq,Choudhury:2023rks,Cheng:2023ikq,Firouzjahi:2023ahg,Davies:2023hhn}.

In this paper, we derive the peak amplitude and the frequency of the PBH-production-induced stochastic gravitational waves (GW) and
estimate quantum loop corrections in the model of Ref.~\cite{Saburov:2023buy} by using the $\delta N$ formalism 
\cite{Cai:2018dkf,Abolhasani:2019cqw}. We use the natural units with $c=\hbar=M_P=1$ where $M_P$ is the reduced Planck mass
in our equations, while restoring them for the values of dimensional physical observables.

\section{The model}

The phenomenological model \cite{Saburov:2023buy} of inflation and PBH production has the $F(R)$ gravity action
\be \lb{MGaction}
 S = \fracmm{1}{2} \int d^4x \sqrt{-g} \,F(R)~~,
\ee
whose $F$-function of the spacetime scalar curvature $R$ reads
\begin{equation} \lb{Ff}
	F(R)=(1+g\tanh b) R + gE_{AB}  \ln \left[\fracmm{\cosh \left(\frac{R}{E_{AB}}-b\right)}{\cosh (b)}\right]
	+\fracmm{R^2}{6 M^2} - \d\fracmm{R^4}{48M^6}~~,
\end{equation}
where the first three terms are known in the literature as the Appleby-Battye-Starobinsky (ABS) model \cite{Appleby:2009uf} with 
the Starobinsky mass $M\approx 1.3\times 10^{-5}$ defining the scale of the first SR phase of inflation.The ABS parameter 
\begin{equation} \lb{ABp}
	E_{AB}=\fracmm{R_0}{2g\ln(1+e^{2b})}
\end{equation}
has the new scale $R_0$ defining the second SR phase of inflation below the Starobinsky scale.~\footnote{It differs from Ref.~\cite{Appleby:2009uf} where $R_0$ was related to the dark energy scale.}  The other parameters $g$ and $b$ define the shape of the inflaton potential and have to be fine-tuned in order to get a near-inflection point. The last term in Eq.~(\ref{Ff}) may be considered as a quantum gravity
correction that was employed in Ref.~\cite{Saburov:2023buy} in order to get good (within $1\sigma$) agreement 
 with the measured CMB value of $n_s$. The function (\ref{Ff}) obeys the no-ghost (stability) conditions, $F'(R)>0$ and $F''(R) >0$, for the relevant values of $R$, avoids singularities, obeys the Newtonian limit and describes double inflation with three phases 
(SR-USR-SR) after fine-tuning the parameters \cite{Saburov:2023buy}.

To produce PBH, one needs a large enhancement of the power spectrum of scalar perturbations by seven orders of magnitude against the CMB spectrum. Then, as was shown in Ref.~\cite{Saburov:2023buy}, the parameters $(R_0,g,b)$ should be fine-tuned as 
\be \lb{parv}
R_0\approx 3.00M^2~,\quad g\approx 2.25 \quad {\rm and} \quad b\approx 2.89~.
\ee
It leads to production of PBH with asteroid-size masses in the range between $10^{16}$ g and $10^{20}$ g \cite{Saburov:2023buy}, exceeding the Hawking (black hole) evaporation limit of $10^{15}$ g, so that those PBH may form part (or the whole) of dark matter (DM) in the current universe \cite{Carr:2020gox}.

\begin{figure}[h]
\begin{minipage}[h]{0.5\linewidth}
\center{\includegraphics[width=0.8\linewidth]{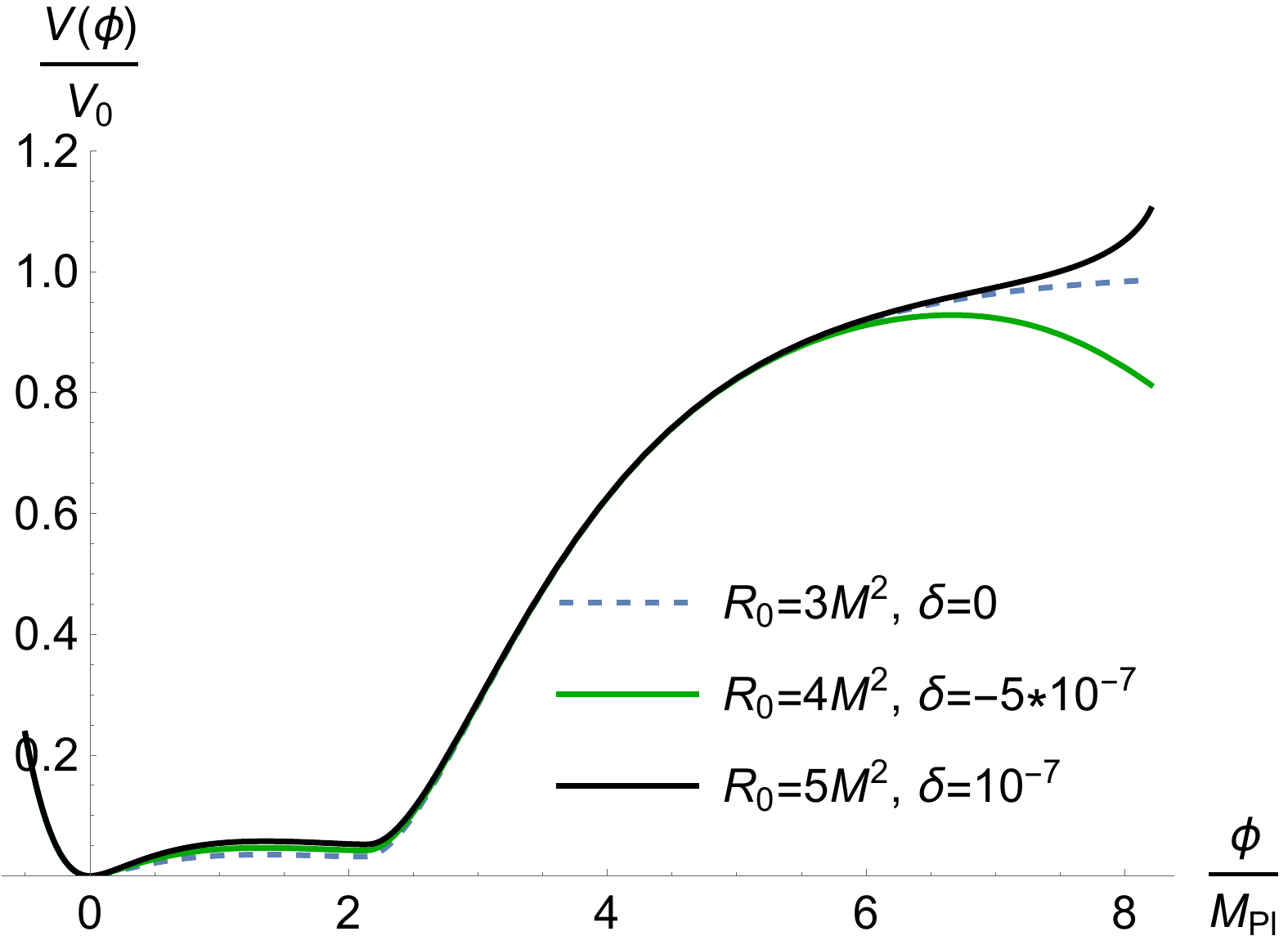} }
\end{minipage}
\hfill
\begin{minipage}[h]{0.5\linewidth}
\center{\includegraphics[width=0.8\linewidth]{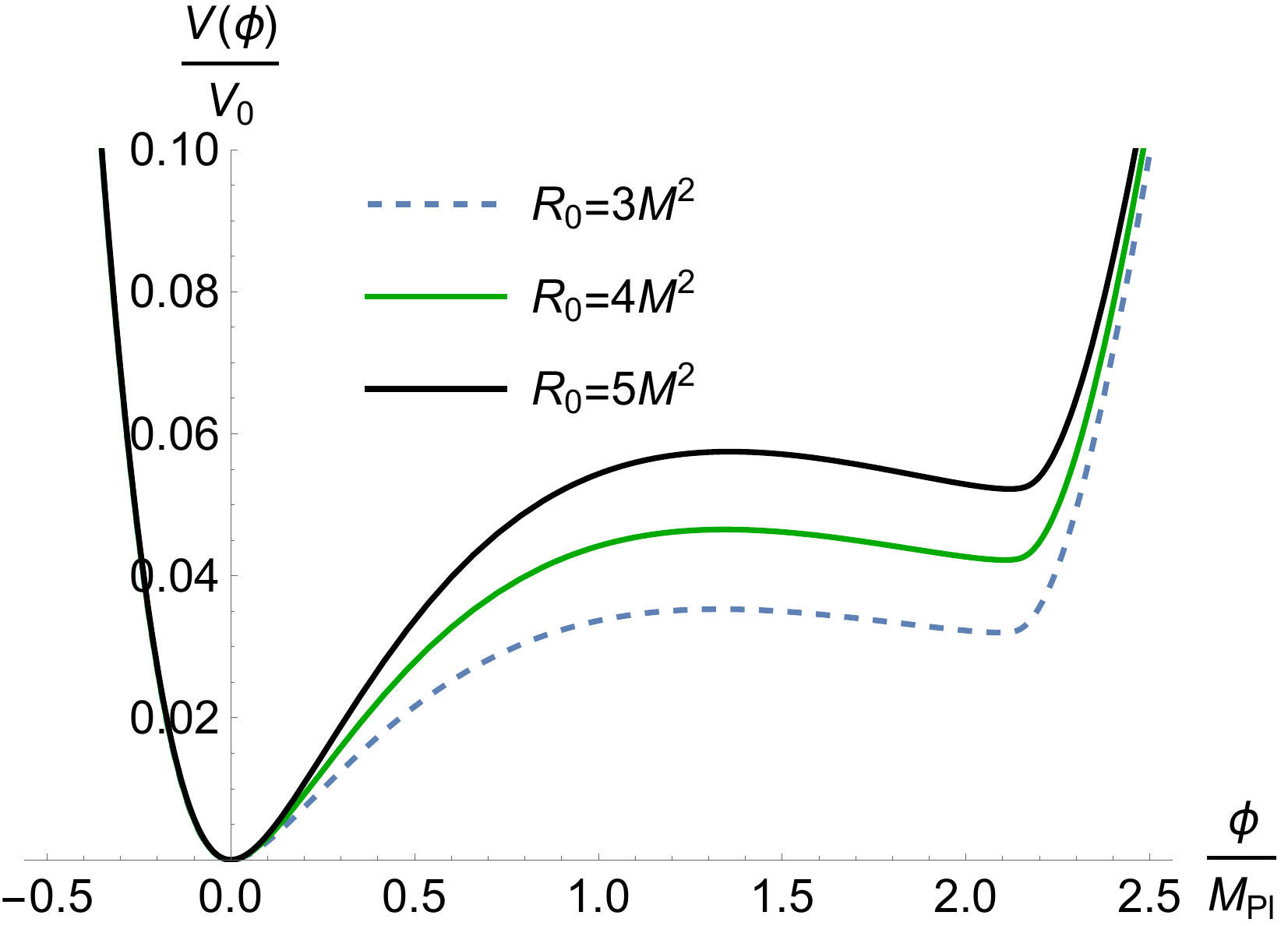} }
\end{minipage}
\caption{\footnotesize {\it On the left:} the inflaton potential having two plateaus for $g=2.25$ and $b=2.89$ with $V_0=\fracmm{3}{4}M^2$. {\it On the right:} zooming the potential for lower values of $\f$ with a near-inflection point. The potential is unstable for negative values of $\d$, and has the infinite plateau for $\d=0$ describing the Starobinsky inflation.}
\label{ris:V1}
\end{figure}

A modified $F(R)$-gravity is known to be equivalent to the quintessence (scalar-tensor gravity) in terms of the canonical inflaton field 
$\phi$ with the scalar potential $V(R(\f))$ in the parametric form \cite{Maeda:1988ab},
\begin{equation}\label{duality}
V(R)=  \fracmm{F'R-F}{2(F')^2}~,\quad	\phi(R)=\sqrt{\fracmm{3}{2}}\ln F'~,
	\end{equation}
where the primes denote the derivatives with respect to $R$. The (numerically obtained) profile of the inflaton potential $V(\f)$ for some values of $R_0$ and $\d$ is given in Fig.~\ref{ris:V1}.

The standard SR conditions are given by $\e_{\rm sr}\ll 1$ and $\abs{\h_{\rm sr}}\ll 1$, where
\be   \lb{sr} \e_{\rm sr} (\f)= \fracmm{1}{2}\left( \fracmm{V'(\f)}{V(\f)}\right)^2\quad {\rm and} \quad 
\h_{\rm sr}(\f) = \fracmm{V''(\f)}{V(\f)}~,
\ee
while the time clock is conveniently defined by the number $N$ of e-folds, $N(t) = \int_{t} H(\tilde{t}) d\tilde{t}$, where $H(t)$ is Hubble function.  The CMB tilt $n_s$ of scalar perturbations and the tensor-to-scalar ratio $r$ are related to the values of the SR parameters at the horizon exit with the standard pivot scale  $k_*=0.05~{\rm Mpc}^{-1}$. In the model \cite{Saburov:2023buy}, the tensor-to-scalar ratio $r$ is well inside the current observational bound, $r<0.032$, and the tilt $n_s$ agrees within $1\s$ with the current CMB measurements~\cite{Planck:2018jri,BICEP:2021xfz,Tristram:2021tvh}, $n_s= 0.9649 \pm 0.0042$, with $\d\sim 10^{-8}$.

\begin{figure}[h]
\center{\includegraphics[width=0.5\linewidth]{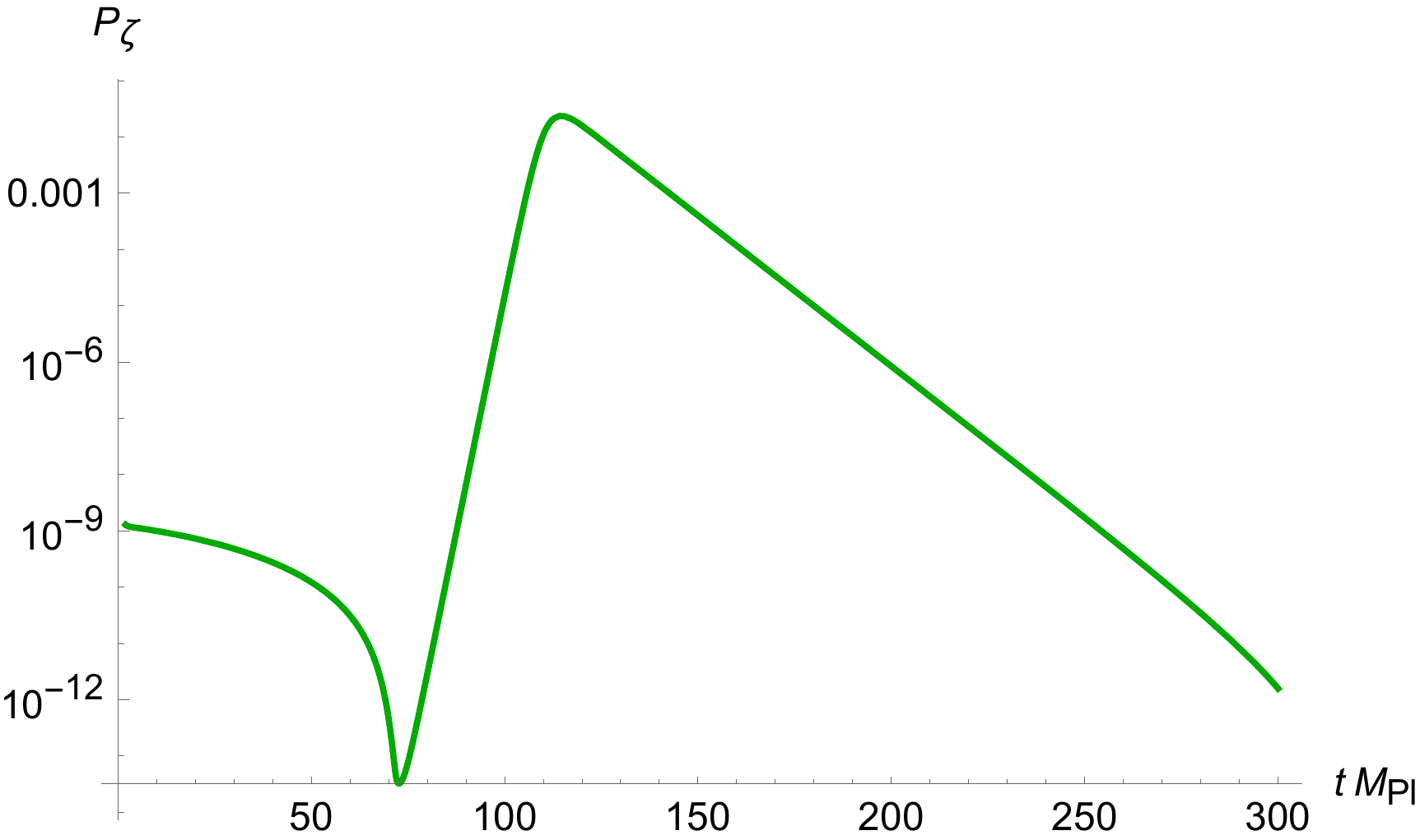}}
\caption{\footnotesize  The primordial power spectrum $P_{\z}(t)$ of scalar perturbations from Eq.~(\ref{powersp}). A derivation of the spectrum from Mukhanov-Sasaki equation leads to the very similar result.}
\label{ris:P0}	
\end{figure}

The primordial spectrum $P_{\z}(k)$ of 3-dimensional scalar (density) perturbations $\z(x)$ in a flat Friedman universe is defined by
the 2-point correlation function as
\be \lb{defpsp}
\VEV{ \fracmm{\d\z(x)}{\z}\cdot \fracmm{\d\z(y)}{\z}} =\int\fracmm{d^3k}{k^3} e^{ik\cdot (x-y)}\fracmm{P_{\z}(k)}{P_0}~~,
\ee
where $k=2\p/\l$ is the co-moving number. The scale $k$ is simply related to the e-folds number $N$ via $N=-\int^k d\tilde{k}/\tilde{k}$.
Though the USR phase has dynamics different from the SR one, the dimensionless power spectrum of scalar perturbations in the SR-approximation 
\begin{equation} \lb{powersp}
	P_{\z}=\fracmm{H^2}{8\pi^2\epsilon_{\rm sr}}
\end{equation}
appears to be a good approximation in the USR phase also. The power spectrum is given in Fig.~\ref{ris:P0} in the best case given by the 3rd row of Table 1 in Ref.~\cite{Saburov:2023buy}. Accordingly, the SR parameter $\e_{\rm sr}$ drops to very low values,
indicating the USR phase.

\begin{figure}[h]
\center{\includegraphics[width=0.5\linewidth]{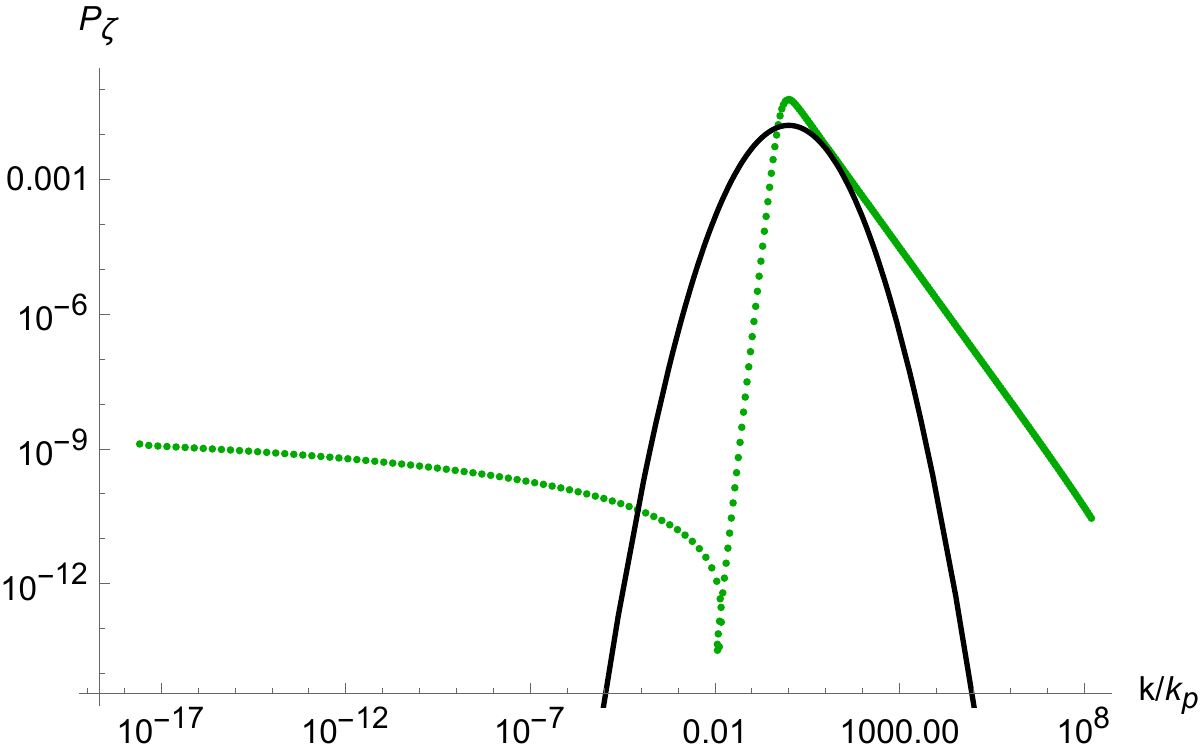}}
\caption{\footnotesize  The primordial power spectrum $P_{\z}(k)$ of scalar perturbations (in green) and
the log-normal fit (in black).}
\label{ris:P3}	
\end{figure}

The power-spectrum-related  observables in our model best case are given by~\cite{Saburov:2023buy}
\be \lb{keyobs}
n_s\approx 0.965, \quad r\approx 0.0095, ~\quad M_{\rm PBH}\approx 1.0 \cdot 10^{20}~{\rm g}.
\ee
The corresponding peak in the power-spectrum of Fig.~\ref{ris:P0} can be roughly approximated by the log-normal fit \cite{Frolovsky:2023xid}
\be \lb{lognorps}
P^{\rm peak}_{\z}(k) \approx \fracmm{A_{\z}}{\sqrt{2\p} \D} \exp \left[ \fracmm{-\ln^2(k/k_p)}{2\D^2}\right]
\ee
with the amplitude $A_{\z}\approx 0.06$  and the width $\D\approx 1.5$, where $k_p\approx 4.5\cdot 10^{12}~{\rm Mpc}^{-1}$ is the location of the peak, see Fig.~\ref{ris:P3}.

\begin{figure}[h]
\begin{minipage}[h]{0.5\linewidth}
\center{\includegraphics[width=0.8\linewidth]{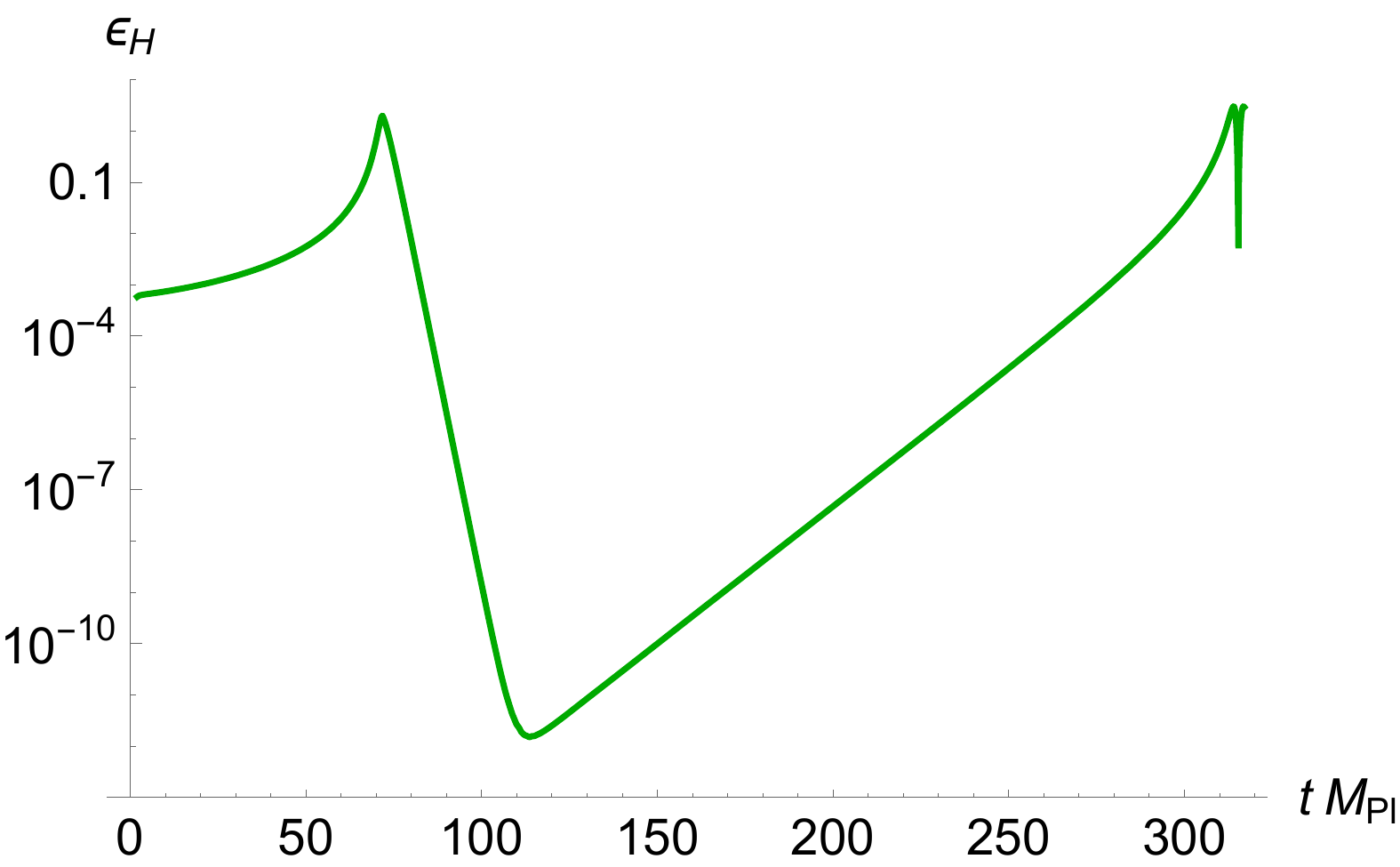} }
\end{minipage}
\hfill
\begin{minipage}[h]{0.5\linewidth}
\center{\includegraphics[width=0.8\linewidth]{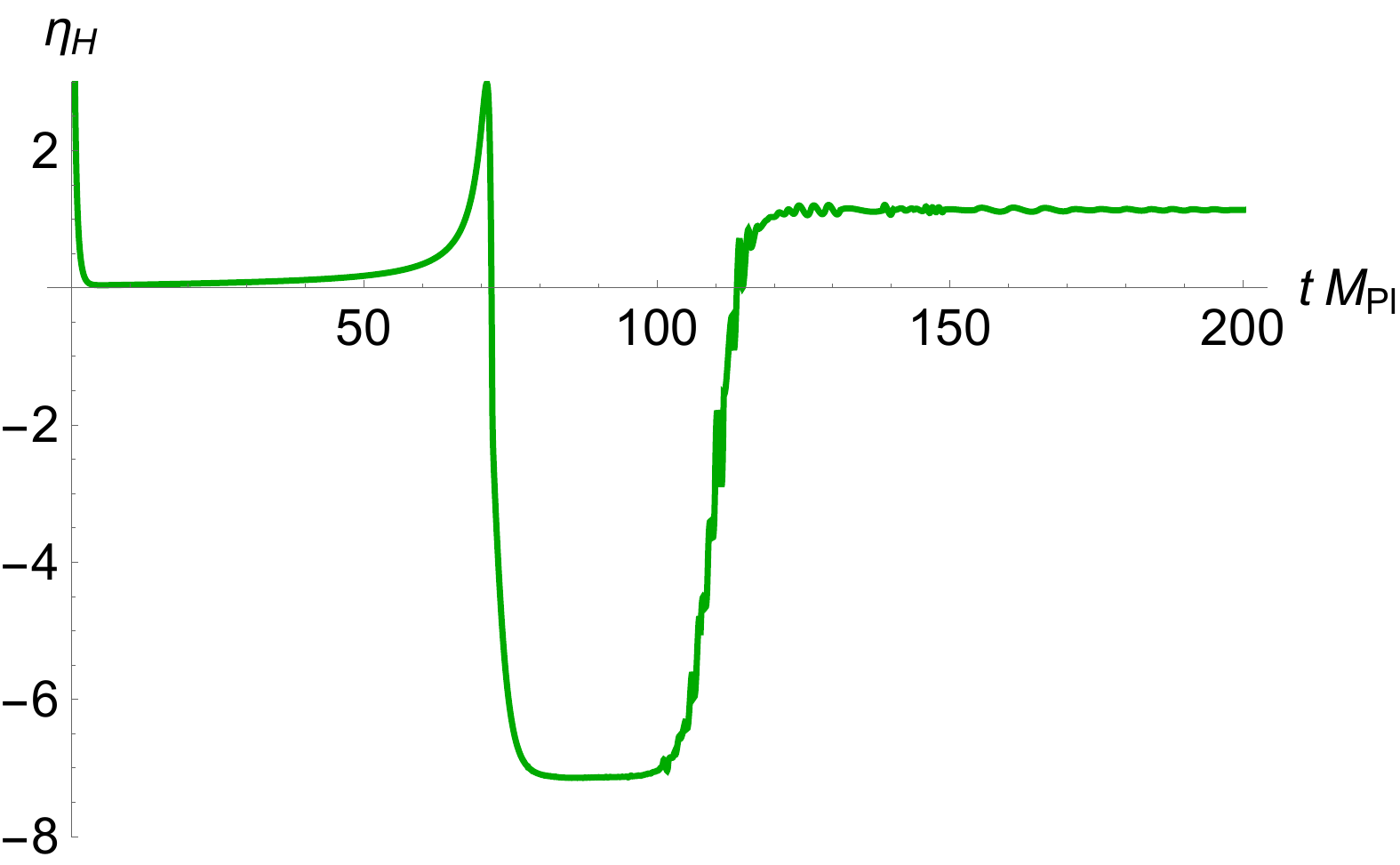} }
\end{minipage}
\caption{\footnotesize The evolution of $\e_H(t)$ and $\h_H(t)$ with the initial conditions  
$\f_{in}=7.01$ and $\dot{\f}_{in}=0$, and the parameters $\d=2.7\cdot 10^{-8}$ and $R_0=3.0~M^2$. In the USR phase, the value of $\h_H(t)$ is close to $-6$.}
\label{ris:Hflow}
\end{figure}

The value of $\h_{\rm sr}\approx -0.025$ in the USR phase practically does not depend upon the parameters $R_0$ and $\d$. The more illuminating functions are given by the Hubble flow parameters 
\begin{equation}\label{Hflow}
	\epsilon_{H} = -\fracmm{\dot{H}}{H^{2}}~,\quad 
	\eta_{H} = \epsilon_{H} - \fracmm{\dot{\epsilon}_{H}}{2\epsilon_{H} H}~~.
\end{equation}
Though the $\e_{\rm sr}$ and $\epsilon_{H}$ can be identified, evolution of $\h_{\rm sr}$ and $\eta_{H}$ is different during the USR phase, see Fig.~\ref{ris:Hflow}. 

\section{PBH-induced GW and PBH-DM density fraction}

PBH production in the early universe leads to stochastic gravitational waves (GW) different from primordial GW caused by inflation. The current energy density  fraction of those PBH-induced GW can be computed in the second order with respect to perturbations as \cite{Espinosa:2018eve,Kohri:2018awv}
\begin{align}
  \Omega_{\mathrm{GW}}(k)=&\fracmm{c_g\Omega_{r,0}}{36}\int^{\fracmm{1}{\sqrt{3}}}_{0}  \,\mathrm{d}d \int^{\infty}_{\fracmm{1}{\sqrt{3}}}  \,\mathrm{d}s \left[\fracmm{(d^2-1/3)(s^2-1/3)}{s^2-d^2}\right]^2\times\nonumber\\
  &\mathcal{P}_\zeta\left(\fracmm{k\sqrt{3}}{2}(s+d)\right)\mathcal{P}_\zeta\left(\fracmm{k\sqrt{3}}{2}(s-d)\right)\left[
  \mathcal{I}_c(d,s)^2+\mathcal{I}_s(d,s)^2\right]~,\lb{eqp}
\end{align}
where the functions $\mathcal{I}_c(d,s)$ and $\mathcal{I}_s(d,s)$ are 
\begin{align}
  \mathcal{I}_c(d,s)&=-36\pi\fracmm{(s^2+d^2-2)^2}{(s^2-d^2)^3}\theta(s-1)~,\nonumber\\
  \mathcal{I}_s(d,s)&=-36\fracmm{s^2+d^2-2}{(s^2-d^2)^2}\left[\fracmm{s^2+d^2-2}{(s^2-d^2)}\mathrm{ln}
  \abs{ \fracmm{d^2-1}{s^2-1}}+2\right]~,
\end{align}
with $\Omega_{r,0}\sim8.6\times10^{-5}$ being the current energy density fraction of radiation, $\theta(s-1)$ is the step function, and $c_g\approx 0.4$.

A numerical calculation of Eq.~(\ref{eqp}) with the power spectrum in Fig.~\ref{ris:P0} yields the result given in Fig.~\ref{gwfit} in green.
The numerical plot can be well approximated by the lognormal fit given in Fig.~\ref{gwfit} in black, with the analytic formula
\be \lb{logngw}
  \O_{\rm GW}(k)=\fracmm{A_{\rm GW}}{\sqrt{2\pi}\sigma_{\rm GW}}\mathrm{exp}\left[-\fracmm{\mathrm{ln}^2(k/k_p)}{2\sigma^2_{\rm GW}}\right]~,
\ee
the amplitude $A_{\rm GW}\approx 5.6\cdot 10^{-8}$ and the width $\sigma_{\rm GW}\approx \D/ \sqrt{2}\approx 1.06$,
where $\D$ is the width of the power spectrum in Fig.~\ref{ris:P0}. The $\Omega_{\mathrm{GW}}^{\rm peak}(k)$ near the peak is roughly given by
$10^{-6}\mathcal{P}^2_\zeta(k)$, in agreement with the estimates in Refs.~\cite{Pi:2020otn,Domenech:2021ztg}. 

\begin{figure}[h]
  \centering
  \begin{minipage}[t]{0.45\hsize}
\center{\includegraphics[width=0.95\linewidth]{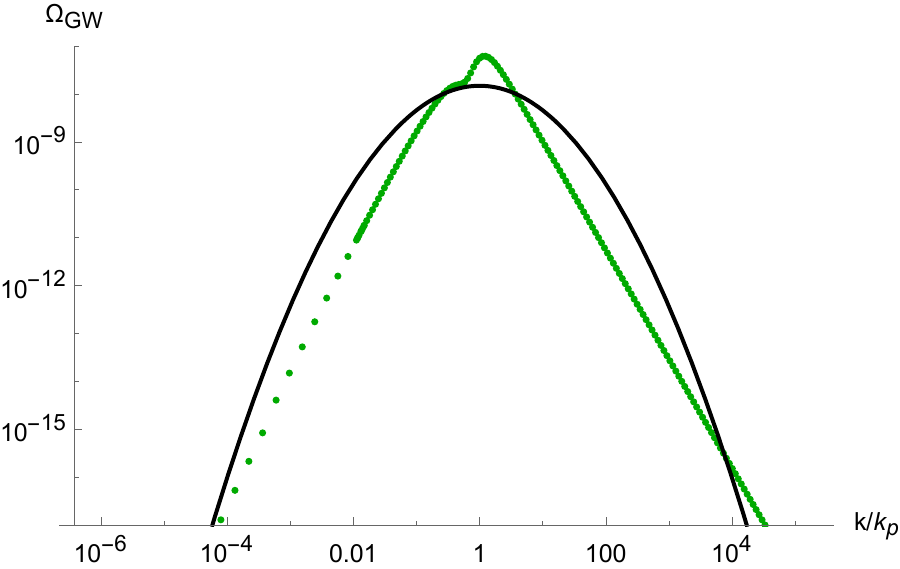}}
  \end{minipage}
  \caption{The PBH-induced stochastic GW density fraction (in green) with the log-normal fit (in black) in Eq.~(\ref{logngw}).}
  \lb{gwfit}
\end{figure}

The induced GW frequencies $f_p$ are related to the PBH masses as \cite{DeLuca:2020agl}
\be \lb{gwf}
  f_p\approx5.7\left(\fracmm{M_\odot}{M_{\mathrm{PBH}}}\right)^{1/2}10^{-9}~\mathrm{Hz},
\ee
where the Sun mass is given by $M_\odot\approx 2\cdot 10^{33}$ g. Given the PBH masses of $10^{20}$ g, as in our model, it results in the GW frequency $f_p\approx 0.0255$~Hz. It is higher than the GW frequencies between 3 and 400 nHz  detected by NANOGrav \cite{NANOGrav:2023gor}. A detection of the GW peak with that frequency in the $10^{-2}$ Hz range would provide observational support to our model provided the peak is due to generation of the secondary GW. A more specific comparison of our predictions with future GW observations is possible by plotting the GW spectrum in our model against the expected sensitivity curves in the future space-based gravitational interferometers such as LISA \cite{LISA:2017pwj,Smith:2019wny}, TianQin \cite{TianQin:2015yph}, Taiji \cite{Gong:2014mca,Ruan:2018tsw} and DECIGO \cite{Kudoh:2005as}, see Fig.~\ref{sensi3}, where we have used Refs.~\cite{Thrane:2013oya,Schmitz:2020syl,Aldabergenov:2020yok} for the colored curves. The  space-based experiments are expected to be sensitive to stochastic GW in the frequencies between $10^{-3}$ and $10^{-1}$ Hz, while the predicted black curve in Fig.~\ref{sensi3} in our model belongs to that frequency range.

\begin{figure}[h]
  \centering
  \begin{minipage}[t]{0.45\hsize}
    \includegraphics[keepaspectratio, scale=1.0]{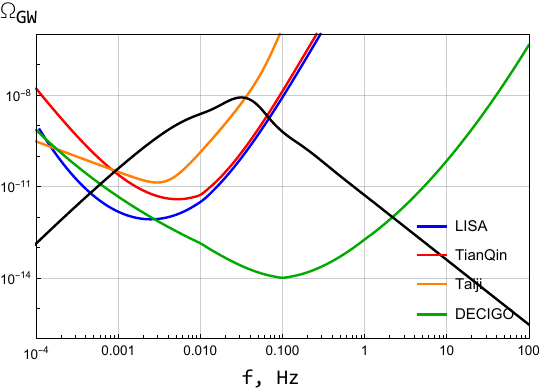}
  \end{minipage}
  \caption{The GW density induced by the power spectrum of scalar perturbations in our model (in black) against the expected sensitivity curves for the future space-based GW experiments (in color).}
\lb{sensi3}
\end{figure}

The PBH-in-DM density fraction $f(k)$ on scale $k$ can be estimated in the Press-Schechter formalism \cite{Press:1973iz} as
\be
    f(k)=\fracmm{\Omega_{\rm PBH}(k)}{\Omega_{\rm DM}}\approx 
    \fracmm{1.2\cdot 10^{24}\beta_f(k)}{\sqrt{M_{\rm PBH}(k){\rm g}^{-1}}}
    \approx 17.1 (k\cdot{\rm Mpc})\beta_f(k) \label{f_PBH}~~,
\ee
where \cite{Inomata:2017okj,Aldabergenov:2020bpt}
\be
\beta_f(k)\simeq\fracmm{\sigma(k)}{\sqrt{2\pi}\delta_c}
    e^{-\fracmm{\delta^2_c}{2\sigma^2(k)}}~,\quad 
    \sigma^2(k)=\fracmm{16}{81}\int\fracmm{dq}{q}\left(\fracmm{q}{k}\right)^4e^{-q^2/k^2}P_\z(q)~,\label{PBH_productionE}
\ee
with the constant $\delta_c$ depending upon the shape of the PBH peak in the power spectrum and representing the density threshold for PBH formation. The integral in Eq.~(\ref{PBH_productionE}) can be estimated as 
\be \lb{sigmaE}
\sigma^2\approx \fracmm{3.84}{81} P_{\z}^{\rm peak}~,
\ee
where $P_{\z}^{\rm peak}$ is the value of the power spectrum at the PBH peak. Then Eqs.~(\ref{f_PBH})  and (\ref{PBH_productionE}) imply 
\be \lb{fracE}
f(k)\sim \fracmm{10^{23}}{\sqrt{M_{\rm PBH}(k){\rm g}^{-1}}}\left(\fracmm{\sqrt{P_{\z}^{\rm peak}}}{\d_c}\right)
e^{-9.3\,\d_c^2/P_{\z}^{\rm peak}}~.
\ee
This equation demonstrates high sensitivity of the PBH-in-DM fraction upon the value of $\d^2_c/P_{\z}^{\rm peak}$. In the
case of the power spectrum in Fig.~\ref{ris:P0}, we have $\d_c\approx 0.45$ and $P_{\z}^{\rm peak}\approx 0.05$. Given the range of the model parameters in Ref.~\cite{Saburov:2023buy}, Eq.~({\ref{fracE}) gives  the PBH fraction in DM between $0.1\%$ and $100\%$. 
In addition, the Press-Schechter formalism itself should be considered with a grain of salt because it was found to be unreliable~\cite{Franciolini:2018vbk,Germani:2023ojx,Figueroa:2021zah}. It is also worth mentioning that even a small PBH fraction could have an important role in cosmology \cite{Carr:2020gox}.

\section{Loop corrections}

In the $\d N$ formalism \cite{Abolhasani:2019cqw} for single-field inflation, a scalar (comoving curvature) perturbation 
$\d N\equiv \z$ is a function of variation $\d \phi$ of inflaton $\phi_i$ at its initial value,
\begin{equation}\label{kor}
	\delta N = N'(\phi)\delta\phi + \fracmm{N''(\phi)}{2!}\delta\phi^{2} + \fracmm{N'''(\phi)}{3!}\delta\phi^{3}+\ldots~,
\end{equation}
where perturbations are not assumed to be small. The power spectrum of scalar perturbations is defined by a two-point function of Fourier components $\z_{\bf p}$ as
\begin{equation}\label{fur}
	\langle\zeta_{\bf p_{1}}\zeta_{\bf p_{2}}\rangle =(2\pi)^{3} \delta^{3}({\bf{p_{1}}}+{\bf p_{2}})P(p)~,\quad
	P_{\zeta}(p)=\fracmm{k^3}{2\p^2}P(p)~,
\end{equation}
where 
\begin{equation}\label{korf}
	\zeta_{\bf p} = N'\delta\phi_{\bf p} + \fracmm{N''}{2!}\int\fracmm{d^{3}q}{(2\pi)^{3}}\delta\phi_{\bf q}\delta\phi_{\bf p-q}+ \fracmm{N'''}{3!}\int\fracmm{d^{3}q_{1}}{(2\pi)^{3}}\fracmm{d^{3}q_{2}}{(2\pi)^{3}}\delta\phi_{\bf q_{1}}\delta\phi_{\bf q_{2}}\delta\phi_{\bf p- q_{1} -q_{2}} + \ldots
\end{equation}
in terms of external 3D momenta ${\bf p}$ and loop momenta ${\bf q}$. Substituting Eq.~({\ref{korf}) into Eq.~(\ref{fur}) yields the loop expansion of the power spectrum $P_{\zeta}(p)$. In order to apply that to a particular model, one has to know the function $N(\phi)$ explicitly. It was derived in Ref.~\cite{Firouzjahi:2023ahg},
\begin{equation}\label{N1}
	N_{\rm total}(\phi_{i}) \simeq \fracmm{1}{\eta_{\rm sr}} \ln\Big[1+\fracmm{\eta_{\rm sr}}{\sqrt{2\epsilon_{\rm sr}}}(\phi_{i} - \phi_{s})\Big] - \fracmm{1}{3} \ln\Big[1+\fracmm{3(\phi_{s} - \phi_{e})}{\pi_{s}}\Big] + \fracmm{1}{\eta_{\rm V}} \ln[-2\eta_{\rm V}\pi_{e} - 6\sqrt{2\epsilon_{\rm V}}]~,
\end{equation}
where the first term refers to the SR(I) phase with the initial value $\phi_i$ and the end value $\phi_s$, the second term refers to the USR phase with the initial momentum value $\pi_s$ and the end momentum value $\pi_e$, and the third terms refers to the SR(II) phase with the slow-roll parameters $\e_{\rm V}$ and $\eta_{\rm V}$. The subscripts $(s,e)$ refer to values of any quantity at the start and end of the USR phase, respectively. The leading contribution comes from the first term in Eq.~(\ref{N1}).

To compute loop corrections, one has to calculate the derivatives $N^{(n)}(\phi)$. The first three derivatives can be estimated as follows:
\begin{equation}\label{n}
	N' \approx \fracmm{1}{\sqrt{2\epsilon_{s}}}\approx\fracmm{e^{-3\Delta N_{\rm USR}}}{\sqrt{2\epsilon_{e}}}~,\quad
	N'' \approx -\fracmm{\eta_{\rm sr}}{2\epsilon_{s}}~,\quad 
	N''' \approx \fracmm{2\eta_{\rm sr}^{2}}{(2\epsilon_{s})^{3/2}}~~,
\end{equation}
where $\Delta N_{\rm USR}$ is the duration of the USR phase and $\eta_{\rm sr}\equiv\eta_{\rm sr}(\phi_{s})$. 

To compute loop corrections to the amplitude of the power spectrum, we considered the effective action up to the third order with respect to $\delta\phi(\bm{x},t)$ on the background $\phi(t)$,
	\begin{equation}
	S_{\delta\phi} = S^{(2)}_{\delta\phi} + S^{(3)}_{\delta\phi} =\fracm{1}{2} \int d^4x \sqrt{-g} \Big(g_{\mu\nu}\partial^{\mu}\delta\phi\partial^{\nu}\delta\phi-V_{,\phi\phi}\delta\phi^{2}\Big) + 
	\fracm{1}{2} \int d^4x \sqrt{-g} \Big(- \fracm{1}{3}V_{,\phi\phi\phi}\delta\phi^{3}\Big).
\end{equation}
A comparable contribution of the quartic coupling  $V_{,\phi\phi\phi\phi}$ was investigated in Ref.~\cite{Firouzjahi:2023aum}.  When using the FLRW background, the effective action reads
\begin{equation}\label{ac}
	S_{\delta\phi} = \fracm{1}{2} \int d^4x \, a^{3}(t)\Big[\dot{\delta\phi}^{2}-a^{-2}(t)(\partial\delta\phi)^{2} - V_{,\phi\phi}\delta\phi^{2}\Big] + \int d^4x\Big[-\fracm{a^{3}(t)}{3!}V_{,\phi\phi\phi}\delta\phi^{3}\Big]~,
\end{equation}
where $\partial\equiv \sum_{i}\partial_{i}$ is a sum of spatial derivatives. The mode functions arising in the solutions to classical equations of motion from the action (\ref{ac})
with Bunch-Davies initial conditions,
\begin{equation}\label{modes}
	u_{k}(\tau) = \fracmm{H}{\sqrt{2k^{3}}} (1+ik\tau)e^{-ik\tau}~,
\end{equation}
are written down in terms of the conformal time $d\tau = \fracm{dt}{a}$, where $-\infty<\tau\leq0$. The CMB modes that left the horizon during inflation are given by $u_{k}(0) = \fracmm{H}{\sqrt{2k^{3}}}$~.

Canonical quantization implies a decomposition into positive and negative parts, as well as the commutation relations (in the
interaction picture)
\begin{equation}\label{phi}
				\begin{split}
	&\delta\phi^{I}_{\bf k}(\tau) = \delta\phi^{+}_{\bf k}(\tau) + \delta\phi^{-}_{\bf k}(\tau) = u_{k}(\tau) a_{\bf k} + u_{k}^{*}(\tau) a^{\dagger}_{\bf -k}~,\\	
	&[a_{\bf p}, a^{\dagger}_{\bf q}] = (2\pi)^{3}\delta^{3}(\bm{p} - \bm{q}),~~ [a_{\bf p}, a_{\bf q}] = [a^{\dagger}_{\bf p}, a^{\dagger}_{\bf q}] = 0~.
				\end{split}
\end{equation}

To get the 1-loop correction according to Eqs.~(\ref{fur}) and (\ref{korf}), one has to evaluate the three-point correlation function. For this purpose we applied the {\it in-in} formalism that gives
\begin{equation}\label{corr}
	\langle \delta\phi_{\bf p}\delta\phi_{\bf q}\delta\phi_{\bf -p-q}\rangle = \langle \overline{T}e^{i\int_{t_{0}}^{t}d\widetilde{t}H_{\rm int}} ~\delta\phi^{I}_{\bf p}\delta\phi^{I}_{\bf q}\delta\phi^{I}_{\bf -p-q}~Te^{-i\int_{t_{0}}^{t}d\widetilde{t}H_{\rm int}}\rangle~,
\end{equation}
where $T$ and $\overline{T}$ stand for the time ordering and anti-time ordering respectively, $t_{0}$ and $t$ are the times associated with the subhorizon and superhorizon scales, respectively, and $H_{\rm int}(\tilde{t})$ is the interaction Hamiltonian in the 3rd order,
\begin{equation}\label{Ham}
	H_{\rm int}(\tilde{t}) = \fracm{1}{3!}a^{3}(\tilde{t})V_{,\phi\phi\phi}\int d^3x(\delta\phi^{I}(\bm{x},\tilde{t}))^{3} = \fracmm{1}{3!(2\pi)^{6}}a^{3}(\tilde{t})V_{,\phi\phi\phi}\int d^{3}kd^{3}\lambda\delta\phi^{I}_{\bf k}(\tilde{t})\delta\phi^{I}_{\bm{\lambda}}(\tilde{t})\delta\phi^{I}_{\bf -k-\bm{\lambda}}(\tilde{t})~.
\end{equation}
The standard (Friedmann and Klein-Gordon) equations of motion yield the following asymptotic approximation for the third derivative of the potential in terms of  the Hubble flow parameters (\ref{Hflow}):
\begin{equation}\label{v'''}
	V_{,\phi\phi\phi} \simeq -\fracmm{3H\dot{\eta_{H}}}{2\sqrt{2\epsilon_{H}}}~~.
\end{equation}

Expanding the T-exponentials in Eq.~(\ref{corr}) to the first order with respect to $H_{\rm int}$,  we find
\begin{equation}\label{corr1}
	\begin{split}
		&\langle \delta\phi_{\bf p}(t)\delta\phi_{\bf q}(t)\delta\phi_{\bf -p-q}(t)\rangle \approx -i \int_{t_{0}}^{t}d\widetilde{t}\langle [\delta\phi^{I}_{\bf p}(t)\delta\phi^{I}_{\bf q}(t)\delta\phi^{I}_{\bf -p-q}(t),H^{int}(\tilde{t})]\rangle\\
		&=2{\rm Im}\Bigg(\int_{t_{0}}^{t}d\widetilde{t}\langle\delta\phi^{I}_{\bf p}(t)\delta\phi^{I}_{\bf q}(t)\delta\phi^{I}_{\bf -p-q}(t)H^{int}(\tilde{t})\rangle\Bigg)~.
	\end{split}
\end{equation}
After substituting Eqs.~(\ref{modes}), (\ref{phi}), (\ref{Ham}) and (\ref{v'''}) into Eq.~(\ref{corr1}) and using Wick's theorem, we derived the three-point correlator as follows:
\begin{equation}\label{corr2}
	\begin{split}
	&\langle \delta\phi_{\bf p}\delta\phi_{\bf q}\delta\phi_{\bf -p-q}\rangle =\\
	& =-\fracmm{1}{2(2\pi)^{6}}{\rm Im}\Bigg(\int_{-\infty}^{0}d\widetilde{\tau}a^{3}(\tilde{\tau})\fracmm{H\eta'(\widetilde{\tau})}{\sqrt{2\epsilon}} \int d^{3}kd^{3}\lambda\langle\delta\phi^{+}_{\bf p}\delta\phi^{+}_{\bf q}\delta\phi^{+}_{\bf -p-q}\delta\phi^{-}_{\bf k}(\tilde{\tau})\delta\phi^{-}_{\bm{\lambda}}(\tilde{\tau})\delta\phi^{-}_{\bf -k-\bm{\lambda}}(\tilde{\tau})\rangle\Bigg) \\ 
	&= -\fracmm{\Delta\eta}{2(2\pi)^{6}}{\rm Im}\Bigg(\fracmm{H_{e}a_{e}^{3}}{\sqrt{2\epsilon_{e}}}
	\int d^{3}kd^{3}\lambda\langle\delta\phi^{+}_{\bf p}(\tau)\delta\phi^{+}_{\bf q}(\tau)\delta\phi^{+}_{\bf -p-q}(\tau)\delta\phi^{-}_{\bf k}(\tau_{e})\delta\phi^{-}_{\bm{\lambda}}(\tau_{e})\delta\phi^{-}_{\bf -k-\bm{\lambda}}(\tau_{e})\rangle\Bigg)\\
	&\approx \fracmm{3!(2\pi)^{3}\abs{\Delta\eta}}{2}{\rm Im}\Bigg(\fracmm{H^{4}_{e}a_{e}^{3}}{\sqrt{2\epsilon_{e}}}\fracmm{H^{3}_{0}(1-iq\tau_{e})^{2}}{(2q^{3})^{2}2p^{3}}e^{2iq\tau_{e}}\Bigg)~,
\end{split}
\end{equation}
where the $H(0)=H_{0}$ denoted the Hubble value during the SR(I), the reference time was chosen at $\tau=0$ because we were only interested in the power spectrum on super-horizon scales relevant to CMB, and $p \ll q$. To avoid divergences, the vacuum expectation value was normalized by the volume of the entire system.

Dynamics of the parameter $\eta_{H}$ implies it is essentially constant everywhere except for the moments of a decrease or an increase (corresponding to $\tau_{s}$ and $\tau_{e}$, respectively), while the moment of the increase is particularly significant (see Fig.~\ref{ris:Hflow}). The approximate solution (\ref{N1}) to the equations of motion in our model is smooth as well as the corresponding $\eta_H(t)$ function defined by the second equation (\ref{Hflow}). To simplify our calculations, we employed the derivative of $\eta_{H}$ with respect to the conformal time as the (Dirac) delta function, $\eta_{H}'(\tau)\sim \delta(\tau - \tau_{e})\Delta\eta$, where $\Delta\eta\approx -6$ is the depth of the pit, inside integrations, which corresponds to a sharp transition. Via integration, the delta function fixes the entire integrand at the time $\tau_{e}$ corresponding to the end of the USR stage.

Equations (\ref{kor}), (\ref{fur}) and (\ref{korf}) lead to a recovery of the tree-level contribution (\ref{powersp}) as the leading term in the loop expansion of the power spectrum, as well as the first (1-loop) contribution as follows:
\begin{equation}\label{loop1}
	P^{\rm 1-loop}_{\zeta}(p) \equiv \fracmm{N'N''}{2} \int\fracmm{d^{3}q}{(2\pi)^{3}}\langle \delta\phi_{\bf p}\delta\phi_{\bf q}\delta\phi_{\bf -p-q}\rangle~.
\end{equation}
After substituting Eqs.~(\ref{n}) and (\ref{corr2}) into Eq.~(\ref{loop1}) we found
\begin{equation}\label{l1}
		\begin{split}
	&P^{\rm 1-loop}_{\zeta}(p) \approx \fracmm{\eta_{\rm sr}e^{-3\Delta N_{\rm USR}}}{2\sqrt{2\epsilon_{e}}2\epsilon_{s}}
	 \int_{\rm USR}dq(4\pi q^{2})\fracmm{3!\abs{\Delta\eta}}{2}{\rm Im}\Bigg(\fracmm{H^{4}_{e}a_{e}^{3}}{\sqrt{2\epsilon_{e}}}\fracmm{H^{3}_{0}(1-iq\tau_{e})^{2}}{(2q^{3})^{2}2p^{3}}e^{2iq\tau_{e}}\Bigg)\\
	&\approx \eta_{\rm sr}e^{-3\Delta N_{\rm USR}}P_{\zeta}(p)P^{\rm PBH}_{\zeta} \fracmm{(2\pi)^{3}\abs{\Delta\eta}}{4}\Bigg(\fracmm{H_{0}}{H_{e}}\Bigg)~,
\end{split}
\end{equation}
where $P^{\rm PBH}_{\zeta}\sim 10^{-2}$ is the fixed amplitude of the power spectrum on the small scales associated with the short-wavelength PBH modes exiting the horizon during the USR phase of inflation. The value of $H_{0}/H_{e}\approx 5$ defines
the ratio of the inflation and PBH scales in our model.

It is evident from Eq.~(\ref{l1}) that the dependence of the 1-loop correction upon the slow-roll parameter $\eta_{sr}$ comes from the second derivative $N''$, the exponential factor depending upon $\Delta N_{\rm USR}$ arises from the first derivative $N'$, and the dependence upon ${P}^{\rm PBH}_{\zeta}$ and $\Delta\eta$ comes from the third derivative $V_{,\phi\phi\phi}$, namely, from the $\eta'(\tau)$.

A detailed calculation of the higher-loop corrections is highly involved and will not be given here. However, it is possible to get
a rough estimate of the 2-loop correction by using the approximative formula given in Ref.~\cite{Firouzjahi:2023ahg},
\begin{equation}\label{loop2}
	P^{\rm 2-loop}_{\zeta}(p) \approx \fracmm{N'N'''}{3!} |\delta\phi_{\bf p}|^{2}\int\fracmm{d^{3}q}{(2\pi)^{3}}|\delta\phi_{\bf q}|^{2} 
	\sim \eta^{2}_{\rm sr}\Delta N_{\rm USR}P_{\zeta}(p)P^{\rm PBH}_{\zeta} ~.
\end{equation}
In our model, according to the plot on the right-hand-side of Fig.~\ref{ris:Hflow}, we have $\Delta N_{\rm USR} \approx 3.1$. 

Therefore, the relative size of the 1-loop and 2-loop corrections from PBH production to the power spectrum at the CMB pivot scale $k^*=0.05$ are
\begin{equation}\label{loop111}
	\fracmm{P_{\zeta}^{\rm 1-loop}(k^*=0.05)}{P_{\zeta}(k^*=0.05)}\approx \eta_{\rm sr}e^{-3\Delta N_{\rm USR}}P^{\rm PBH}_{\zeta} \fracmm{(2\pi)^{3}\abs{\Delta\eta}}{4}\Bigg(\fracmm{H_{0}}{H_{e}}\Bigg)\approx 10^{-3}
\end{equation}
and
\begin{equation}\label{loop222}
	\fracmm{P_{\zeta}^{\rm 2-loop}(k^*=0.05)}{P_{\zeta}(k^*=0.05)}\sim \eta^{2}_{\rm sr}\Delta N_{\rm USR}P^{\rm PBH}_{\zeta}\approx10^{-5}~,
\end{equation}
where we have used $P_{\zeta}(k^*=0.05)\approx2\cdot10^{-9}$ for the CMB power spectrum. Therefore, the one-loop contribution is suppressed by the factor $\eta_{\rm sr}e^{-3\Delta N_{\rm USR}}$, whereas the two-loop contribution is suppressed by $\eta^2_{\rm sr}\Delta N_{\rm USR}$  (we recall that $\eta_{\rm sr}\approx -0.025$ in our model). As regards the higher $n$-loop corrections, their structure includes the suppression factor 
$\eta^n_{\rm sr}P^{\rm PBH}_{\zeta}\sim 10^{-2n-2}$ so that they are expected to be negligible too.

The relative smallness of loop corrections in our model is in agreement with the considerations of Refs.~\cite{Firouzjahi:2023ahg,Davies:2023hhn,Inomata:2024lud,Kristiano:2024ngc,Kristiano:2024vst} but in disagreement with the results of Refs.~\cite{Kristiano:2022maq,Choudhury:2023rks}. Our
calculations were based on the $\delta N$ formalism, also used in 
Ref.~\cite{Firouzjahi:2023ahg}, whereas the calculations performed  in Refs.~\cite{Kristiano:2022maq,Choudhury:2023rks} were based on the {\it in-in} formalism. 
It is beyond the scope of our investigation to compare the two 
formalisms.~\footnote{See, however, Ref.~\cite{Firouzjahi:2023aum} for a partial comparision.}

The amplitude of the power spectrum during USR was fixed in our approach, while we effectively assumed a sharp transition in part of our analytic calculations. The sharpness of transitions can be quantitatively estimated by the parameter $h$ defined by \cite{Cai:2018dkf}
\begin{equation}\label{h}
h = 6\fracmm{\sqrt{2\epsilon_{\rm V}}}{\dt{\phi}(t_e)}=
 -6\fracmm{\sqrt{2\epsilon_{\rm V}}}{\pi_{e}}=
 -6\sqrt{\fracmm{\epsilon_{\rm V}}{\epsilon_s}}e^{3\Delta N_{\rm USR}}~,
\end{equation}
where $\pi_{e}$ is the inflaton momentum at the end of the USR inflation, $\epsilon_{s}\approx 1$ is the SR parameter at the end of the SR(I) or at the beginning of USR, and
$\epsilon_{\rm V}$ is the SR parameter at the end of USR.
In our model, by using Fig.~\ref{ris:Hflow}, we got $h\approx -0.66$ that implies a sharp (though rather mild) transition because, on the one hand,  $\abs{h}$ is not much less than one but, on the other hand, it is still away from a truly sharp transition with the "standard" value  $h= -6$ used in Ref.~\cite{Kristiano:2022maq}. As was demonstrated  in Ref.~\cite{Firouzjahi:2023aum}, the lower value of $h$ also justifies ignoring the quartic coupling in our analysis. As a result, the one-loop correction in our model appeared to be small against the tree-level contribution, as in  Refs.~\cite{Davies:2023hhn,Tada:2023rgp}. 

\section{Conclusion}

The main new results of this paper are given by Fig.~\ref{sensi3}, Eqs.~(\ref{fracE}) and (\ref{l1}).
It follows that the modified gravity model \cite{Saburov:2023buy} of Starobinsky inflation with PBH production may generate a significant part (or the whole) of dark matter from PBH, while it is not ruled out by quantum loop corrections because the latter are relatively small by the factor of $10^{-3}$ against the tree-level (classical) contribution. The key role in the last  conclusion was played by the derivatives of the function $N(\phi)$ during the USR phase, 
describing superhorizon curvature perturbations in the $\delta N$ formalism, which led to the suppression of loop contributions. It is worth mentioning that our results only apply to the particular phenomenological model of PBH production related to Starobinsky inflation.

The predicted frequency $f_p\approx 2.55\cdot 10^{-2}$ Hz of the PBH-production-induced stochastic GW is in the range between $10^{-3}$ Hz and $10^{-1}$ Hz of the frequencies that are expected to be sensitive to the future space-based gravitational interferometers. 

As was recently pointed out in the literature \cite{Dvali:2020wft,Michel:2023ydf,Alexandre:2024nuo,Thoss:2024hsr}, the standard result for the primordial black hole survival at present, based on the Hawking semiclassical evaporation formula, may be relaxed below $10^{15}$ g, when going beyond the semiclassical approximation. Should this be the case, fine-tuning of the parameters in our model for efficient PBH production (needed for DM) may be significantly relaxed.

\section*{Acknowledgements}

We acknowledge discussion and correspondence with Matthew Davies, Gia Dvali, Guillem Domenech, Andrew Gow, Jason Kristiano, Peter Kazinsky, Sayantan Choudhury, Kin-Wang Ng and Alexei Starobinsky.

SS and SVK were supported by Tomsk State University. SVK was also supported by Tokyo Metropolitan University, the Japanese Society for Promotion of Science under the grant No.~22K03624, and the World Premier International Research Center Initiative (MEXT, Japan).

This paper is devoted to memory of late Alexei Starobinsky.

\section*{Data Availability Statement} 
No data associated with the manuscript.

\bibliography{Bibliography}{}
\bibliographystyle{utphys}

\end{document}